\documentclass[journal]{IEEEtran}
\usepackage{graphicx}
\usepackage{caption}
\usepackage{subcaption}
\usepackage{float}
\usepackage{array}
\usepackage{physics}
\usepackage{optidef}
\usepackage{amsmath,amsthm,amsfonts,amssymb,amscd}
\usepackage{xcolor}
\usepackage{algorithm}
\usepackage{algpseudocode}
\usepackage{bm}
\usepackage[numbers, sort&compress]{natbib}

\usepackage{url}
\usepackage[colorlinks=true,linkcolor=black,anchorcolor=black,citecolor=black,filecolor=black,menucolor=black,runcolor=black,urlcolor=black]{hyperref}
\def\BibTeX{{\rm B\kern-.05em{\sc i\kern-.025em b}\kern-.08em
    T\kern-.1667em\lower.7ex\hbox{E}\kern-.125emX}}

\ifCLASSINFOpdf
\else
\fi

\begin{document}
%
\title{Improving Photovoltaic Hosting Capacity of Distribution Networks with Coordinated Inverter Control - A Case Study of the EPRI J1 Feeder}
%
%
%

\author{Dhaval~Dalal,~\IEEEmembership{Senior Member,~IEEE}, Madhura Sondharangalla,~\IEEEmembership{Student Member,~IEEE}, Raja Ayyanar,~\IEEEmembership{Fellow,~IEEE}, and
        Anamitra~Pal,~\IEEEmembership{Senior
        Member,~IEEE}
\thanks{This work was supported in part by the US Department of Energy under grant DE-EE0009355.

The authors are with the School of Electrical, Computer, and Energy Engineering, in Arizona State University (ASU). Email: ddalal2@asu.edu; mbhashit@asu.edu; rayyanar@asu.edu; Anamitra.Pal@asu.edu.
}      
}

%
%

\markboth{}
{Shell \MakeLowercase{\textit{et al.}}: Improving Photovoltaic Hosting Capacity of Distribution Networks with Coordinated Inverter Control - A Case Study of the EPRI J1 Feeder }
%



\maketitle

\begin{abstract}
Adding photovoltaic (PV) systems in distribution networks, while desirable for reducing the carbon footprint, can lead to voltage violations under 
high solar-low load
conditions.
The inability of traditional volt-VAr control in eliminating all the violations is also well-known.
This paper presents a novel coordinated inverter control methodology
that leverages system-wide situational awareness 
to significantly improve
hosting capacity (HC). The methodology employs a real-time voltage-reactive power (VQ) sensitivity matrix in an iterative linear optimizer to calculate the minimum reactive power intervention from PV inverters 
needed for mitigating
over-voltage without resorting to active power curtailing or requiring step voltage regulator setting changes. The algorithm is validated using the EPRI J1 feeder under an extensive set of realistic use cases and is shown to provide 3x improvement in HC under all scenarios.
\end{abstract}


%
\IEEEpeerreviewmaketitle

\section{Introduction}
%
%
%
%
\IEEEPARstart{I}{n} 2022, global solar photovoltaic (PV) generation achieved a record increase of 270 TWh \cite{IEA_1}. However, meeting the net zero emissions (NZE) target of solar PV  generation of 8300 TWh by 2030 (from the current level of 1300 TWh) requires an annual average growth rate of 26\% between 2023 and 2030. In a similar study \cite{osti_1820105}, it was reported that solar PV deployment in the U.S. must quadruple from the current rate of 15 GW/year to 60 GW/year between 2025 and 2030 to meet the NZE target. 
While utility-scale PV additions continue to dominate the capacity (about half of the 2022 capacity additions came from utility-scale) \cite{GlobalSolar},
it is expected that PV additions at the distribution-level
will play an ever-increasing role in achieving the desired growth rate. This is because the
economics and complexities of building utility-scale PV systems are already reaching a saturation phase in many geographies. 
Conversely, distribution-level PV systems
are heavily supported by economic incentives and provide a significant growth opportunity because of a relatively low base.

However, there are a number of challenges that must be overcome to achieve the desired PV proliferation in distribution networks. 
One major challenge is the utilities' limited planning control and visibility over PV deployment in the distribution networks \cite{Ustun}.
Coupled with the inherent variability and uncertainties related to PV generation, it compromises the utilities' ability to provide reliable power under all operating conditions. Furthermore, the standards for distribution-level PV
systems to interface with the grid are still evolving \cite{8332112}. Many of the installed PV inverters have legacy grid-following characteristics, and only a fraction of them have the capability to provide additional grid support services such as voltage support through volt-VAr (VV) or other equivalent control methods \cite{IREC_tracker}. While these limitations are not a major concern at low PV penetration levels (around 10-20\%), they cannot prevent major voltage violations when the PV penetration levels start increasing beyond 30\% \cite{7784836,990181}. 
In summary, the main causes for voltage violations in PV-rich systems are \cite{6096349}:
\begin{itemize}
\item Power flow reversal caused by high PV generation
\item Low X/R ratio of distribution lines
\item Inappropriate regulator and capacitor settings
\item Cross-coupling effects between different phases
\item Sudden drastic changes in PV generation or load profiles
\end{itemize} 

Finally, due to a lack of adequate system-wide situational awareness,
voltage violations caused by the above-mentioned factors are not  quickly detected by the utilities.

Hosting capacity (HC) is a characterization of the level of distributed energy resource (DER) penetration in a network without causing voltage violations or other reliability issues. HC studies abound in literature, but because of the myriad factors involved, they are both complex and diverse in their scope and results \cite{HCGuidebook, MULENGA2021106928}.
The conventional approach for increasing HC of solar PV is to
use the inverter settings to provide reactive power support.
These settings can range from fixed power factor to employing VV control for each inverter \cite{7101709}. Even though these techniques provide some additional HC, they are usually not effective in improving HC beyond 30\%. One alternative for achieving higher HC is to intentionally oversize PVs to allow higher reactive power \cite{Ali}, but such an approach incurs inverter cost penalties and often requires active power curtailment (APC). 
In \cite{osti_1432760}, HC was improved by combining on-load tap-changer (OLTC)/step voltage regulator (SVR) control with local inverter control.
However, as OLTCs/SVRs are mechanical devices, it is better to find a solution that \textit{does not require} their settings to be changed frequently. 
Lastly, \cite{osti_1432760,7101709,Ali} employed uncoordinated voltage control techniques that only relied on the
point of common coupling (PCC)  of the grid-connected DERs for decision-making. 
Even though these control schemes may be
fast, simple, and reliable \cite{9677119}, the limited observability hinders the voltage regulation beyond the PCC, especially with the presence of SVRs and capacitor banks. 

Communication-based methodologies in which the inverters are actively monitored and commanded in a coordinated fashion using information that is not locally available to each inverter have also been proposed
\cite{Abad, Jafari,8954624, 8973721,  Lusis, Yao}. The level of coordination employed varies 
(e.g., comparison of local inverter control with coordinated OLTC/SVR control \cite{Abad};  coordination between global OLTC/SVR control and local inverter control \cite{Jafari, 8954624}; optimization-based coordinated inverter control \cite{8973721, Lusis}; real-time optimal power flow-based coordinated inverter control \cite{Yao}), leading to a diversity of attained benefits.
However, these approaches are either unsuitable for fast time-scale implementation or require significant approximations to achieve real-time implementations.  

A number of recent works have clustered the distribution networks into zones and used machine learning (ML), including self-learning and reinforcement learning, to 
achieve decentralized and faster voltage regulation 
\cite{Zon_1, Zon_2,9805763}. 
Some of them reduce the voltage violations but do not completely eliminate them, while others resort to APC to achieve the required voltage regulation. Moreover, zoning results in a significant barrier to higher improvement in HC due to the \textit{cross-phase effects} as described in this paper (see Section \ref{Cross_Phase}). 
In summary, existing literature has proposed a number of different methodologies to address voltage violations caused by high PV penetration in distribution networks. However, none of them demonstrated full elimination of over-voltage (OV) conditions in a complex distribution network\footnote{A complex distribution network is one that has both primary and secondary circuits, unbalanced multi-phase lines/loads, and SVRs/OLTCs/capacitors.} without resorting to APC and/or changing SVR/OLTC controls. 

In this paper, a novel approach to achieving higher HC in complex distribution networks is implemented that assumes the availability of full voltage information of all the nodes in the system provided by advanced state estimation techniques. The proposed methodology envisages a coordinated controller that uses this information coupled with an iteratively refined sensitivity matrix (SM) in a linear optimization algorithm to provide reactive power (Q) command values for all the DERs in the system. Because of the prioritization embedded in the optimization formulation, this technique results in the activation of a small number of inverters to provide the Q-support while still achieving full voltage mitigation and without requiring APC on any of the PVs. The methodology was validated on a
complex distribution network (the EPRI J1-Feeder) using real-world load and PV profiles. 



The major contributions of this paper are as follows:

\begin{itemize}
\item A novel coordinated inverter control methodology that achieves 3x improvement in HC over the baseline.
\item A real-time voltage-reactive power (VQ) sensitivity matrix implementation that overcomes the limitations of a fixed sensitivity matrix.
\item First (to the authors' knowledge) to incorporate the cross-phase sensitivity effects in voltage control algorithms to achieve better OV mitigation results.
\item Avoidance of any APC while achieving high HC 
across a large range of use cases on a complex feeder.
\item Elimination of the dependence of the control methodology on SVR/OLTC/capacitor setting changes to achieve the required voltage mitigation results.

\end{itemize}


\section{Establishing Baseline HC}\label{Section2}

HC is commonly expressed in percentage as a ratio of the total nameplate wattage capacity of all PV units on the feeder and the peak load (in watts) of the feeder \cite{osti_1432760}.  
The HC depends on a large number of factors related to the distribution networks, as described below.
\begin{itemize}
    \item \textbf{Feeder load characteristics:} Individual loads on any feeder have diurnal and seasonal patterns in addition to a degree of inherent variability. Voltage violations that limit HC typically occur when the load is not at the peak and the PV generation is contemporaneously high. Any HC evaluation should cover a representative range of load profiles in order to establish a robust HC value. 
    \item \textbf{PV characteristics:} The locations of PV systems play an important role in determining HC. If the PVs are concentrated at the far end of the distribution network, HC is severely restricted, and vice-versa \cite{osti_1432760}. The size of the residential PVs is another factor - if the relative PV capacity is much higher than the load, reverse power flow happens, and the chances of voltage violations increase. Finally, the diurnal and seasonal generation patterns of the PVs also play a role. Therefore, using representative patterns derived from diverse real-world datasets help in determining a more realistic HC.  
    \item \textbf{SVRs, OLTCs, Capacitor banks:} 
    These devices are set to provide voltage support based on the time of day/season, load conditions, and distance from the feeder-head. HC calculations should assume that the addition of PVs does not impact the settings of these elements, as they are not designed for frequent switching. 
    \item \textbf{Other feeder characteristics:} Distribution networks are typically unbalanced 3-phase networks. Moreover, they have wide variations in X/R ratios in primary and secondary circuits. These characteristics make determining HC more challenging for distribution systems in comparison to transmission systems.    
\end{itemize}

Thus, any new algorithm for HC enhancement should be validated  on the right distribution network in terms of size, elements, configuration, and data/model access. 

\subsection{EPRI J1 Feeder Characteristics} \label{sub2_1}

\par
\begin{figure}[H]
\centering
  \includegraphics[width=0.48\textwidth]{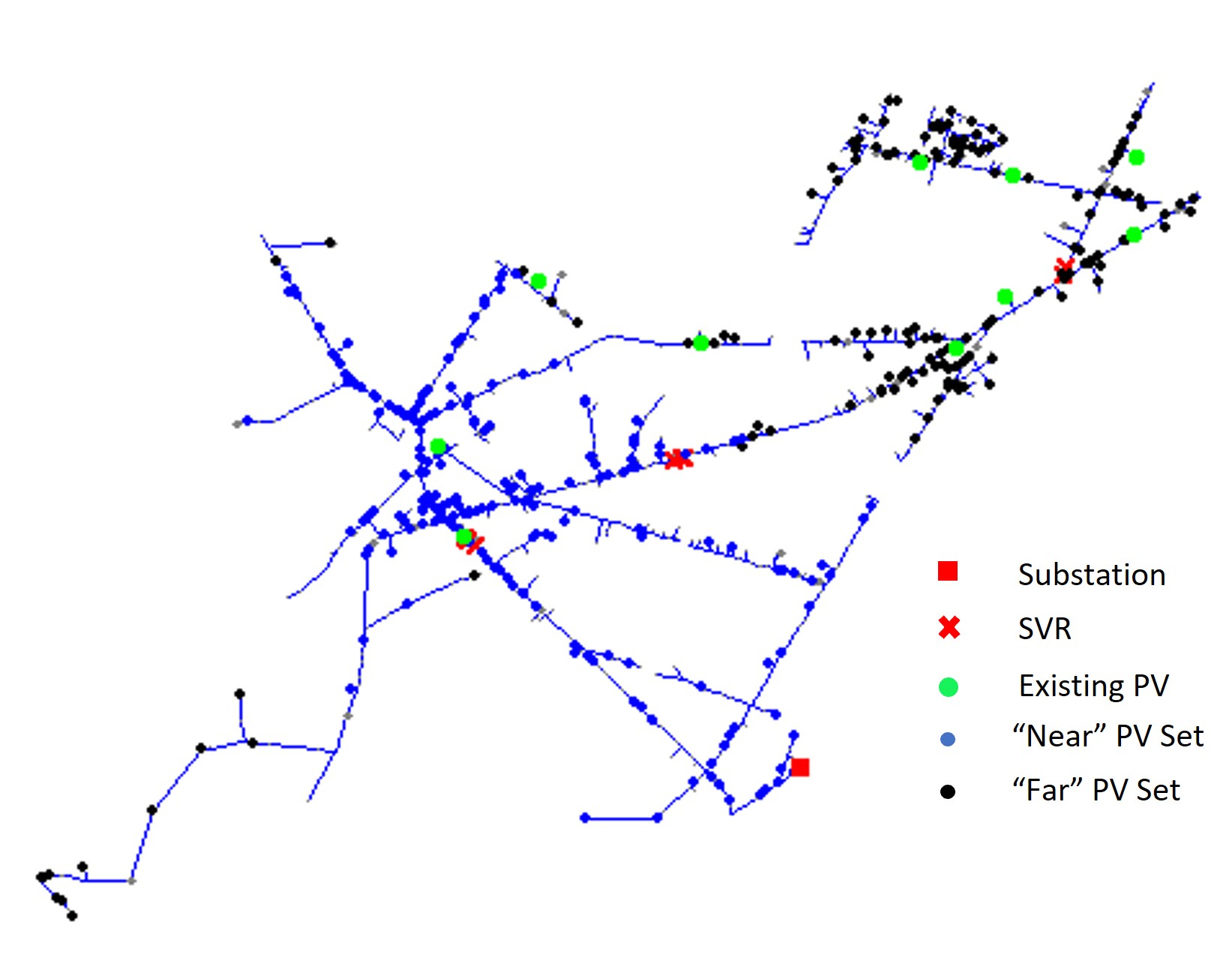}
  \caption{J1 feeder circuit with SVRs and PV additions}
  \label{fig: JCirc}
\end{figure}

 The J1 feeder \cite{en10111896} is a feeder located in the U.S. Northeast with open-source data and models provided by EPRI 
\cite{sourceforgeOpenDSSCode}. 
There have been a number of distribution system studies performed using the J1 feeder \cite{6939028,7101709,7335204} that allow a basis for validation and comparison of the proposed methodology to prior approaches. Key attributes of the J1 feeder are summarized in Table \ref{J1_att}. The following factors make this feeder (see Fig. \ref{fig: JCirc}) suitable for HC studies:
\begin{itemize}
    \item \textit{Appropriate size -} When a network is too small, the HC study can become trivial or guided by one or two dominating factors. Conversely, very large feeders can pose power flow convergence problems when PVs are added to them. In this regard, the J1 feeder presents a very good mid-size option for HC studies.
    \item \textit{Diversity of line characteristics -} The J1 feeder has similar
    number of primary and secondary nodes with detailed line parameters for both primary and secondary networks. These line data have a wide variation that can have an impact on HC. Many other open-source feeders do not have this level of data available and require researchers to make assumptions that may skew the results. 
    \item \textit{Diversity of elements -} The presence of multiple SVRs and capacitors in the J1 feeder represents the level of complexity in the distribution network that any HC improvement algorithm must deal with for demonstrating effectiveness in real situations. 
    Note that the OLTC at the J1 feeder-head is treated as an SVR in this study.
    \item \textit{Diversity of loads and applicable load patterns -} The residential loads in the J1 feeder range from 100s of
    watts to 10s of kilowatts. Even though time series data for the load patterns are not available for the J1 feeder, the fact that the feeder is from the U.S. Northeast makes it easy to apply the load patterns from the Pecan Street data \cite{pecandataport}. 
\end{itemize}

\begin{table}[H]
\caption{Attributes of EPRI J1 feeder}
\label{J1_att}
\centering
\renewcommand{\arraystretch}{1.5} 
\resizebox{\columnwidth}{!}{%
\begin{tabular}{|l|c|l|}
\hline
          \textbf{Parameter}                  & \textbf{Value}    & \textbf{Comments}     \\\hline
Primary Voltage   & 12.47 kV   & Substation transformer at 69 kV         \\\hline
        Secondary Voltage                   & 240 V  &       \\\hline
       Total Customers    & 1384    & 363/375/643/3 (Phase A/B/C/3-ph)         \\\hline
   Total Load   & 10.95 MW & Includes 5 MW aggregated load  \\\hline
   SVRs & 9 & 3/3/2/1 (Phase A/B/C/Substation) \\\hline
   Capacitors & 3900 kVAR & 5 Capacitors \\\hline
   Existing PV & 1813.6 kW & 1.71 MW commercial, 103.6 kW residential \\\hline 
\end{tabular}
}
\end{table}

\subsection{PV Addition Methodology} \label{sub2_2}
The J1 feeder has existing PV generation that includes four 3-phase PV generators totaling 1.71 MW capacity and nine residential PV units adding up to 103.6 kW capacity. For the HC increase, it is more critical to explore the residential PVs as they are more likely to cause voltage violations. Moreover, the existing 1.71 MW commercial PV installation is sufficiently large not to warrant further additions. So, all the new PV additions were made on the residential nodes. The size of the new PVs was standardized to 10 kW $P_{mpp}$ to be consistent with the 
existing residential PV systems of the feeder. 

The location of the added PVs is another important consideration. In order to evaluate the results in the context of previously published results (e.g., \cite{osti_1432760}), three lists were created: one by randomly selecting from all loads on the feeder (``All" set), another by randomly selecting only from the near-half of the feeder (``Near" set), and the third one by randomly selecting from the far-half of the feeder (``Far" set). Randomizing ensures that there is no skewing of the results based on selective PV placements. It is expected that the ``Far" set will result in lower HC than the ``Near" set, while the results from the ``All" set are more representative of the typical PV additions. The PV additions for the ``Near" and the ``Far" sets are marked in different colors in Fig. \ref{fig: JCirc} (blue nodes vs. black nodes) to indicate the cut-off between the two sets. 


\subsection{Application of Pecan Street Data} \label{sub2_3}
One dataset of the 
Pecan Street Project \cite{pecandataport} provides granular data for the electricity consumption and PV generation for New York State. 
This data is available 
for 25 residences (14 of which have PV installed) for a six-month period (May to October). For each residence, representative diurnal profiles for each month were generated based on averaging the data for each hour. These profiles were normalized, and five load profiles and six PV profiles were selected for application to the J1 feeder loads and PV systems. The application of these profiles was randomized within the J1 system. 
This methodology ensured that the HC evaluations were based on representative real-world load and PV data and also incorporated the diversity normally present in such feeders. 




\subsection{Capacitor and SVR Settings} \label{sub2_4}
Typically, settings of SVRs and capacitor banks are changed infrequently (often on a seasonal basis). 
This is because both of them involve mechanical switching, and hence, their frequent usage can lead to significant wear and tear (and associated costs). In other words, they are not designed to react to diurnal PV variations.
At the same time, these devices can 
have an adverse effect on HC based on their control parameters and values.
The approach taken to determine the capacitor and SVR levels and settings for this study is as follows:
\begin{enumerate}
    \item For a given month, apply the load and PV profiles from Pecan Street data to the J1 feeder as per Section \ref{sub2_3}.
    \item Select initial capacitor values and control settings and SVR tap settings. 
    \item Run daily mode power flows in a distribution system solver (such as OpenDSS \cite{OpenDSS}).
    \item Identify any voltage violations (typically OV during early AM hours and/or under-voltage (UV) during late PM hours). Note that the PV systems are not contributors to these violations - they are primarily artifacts of the load and capacitor/SVR settings.
    \item Empirically adjust the capacitor and SVR settings till these violations are minimized in number and values. Note that this is a balancing act between two conditions - increasing the tap settings helps avoid UV violations while potentially increasing OV violations. This process is iterative and does not necessarily result in zero violations under all conditions. 
    \end{enumerate}

\begin{figure}
\centering
\begin{subfigure}[b] {0.48\textwidth}
    \centering
    \includegraphics[width=1\linewidth]{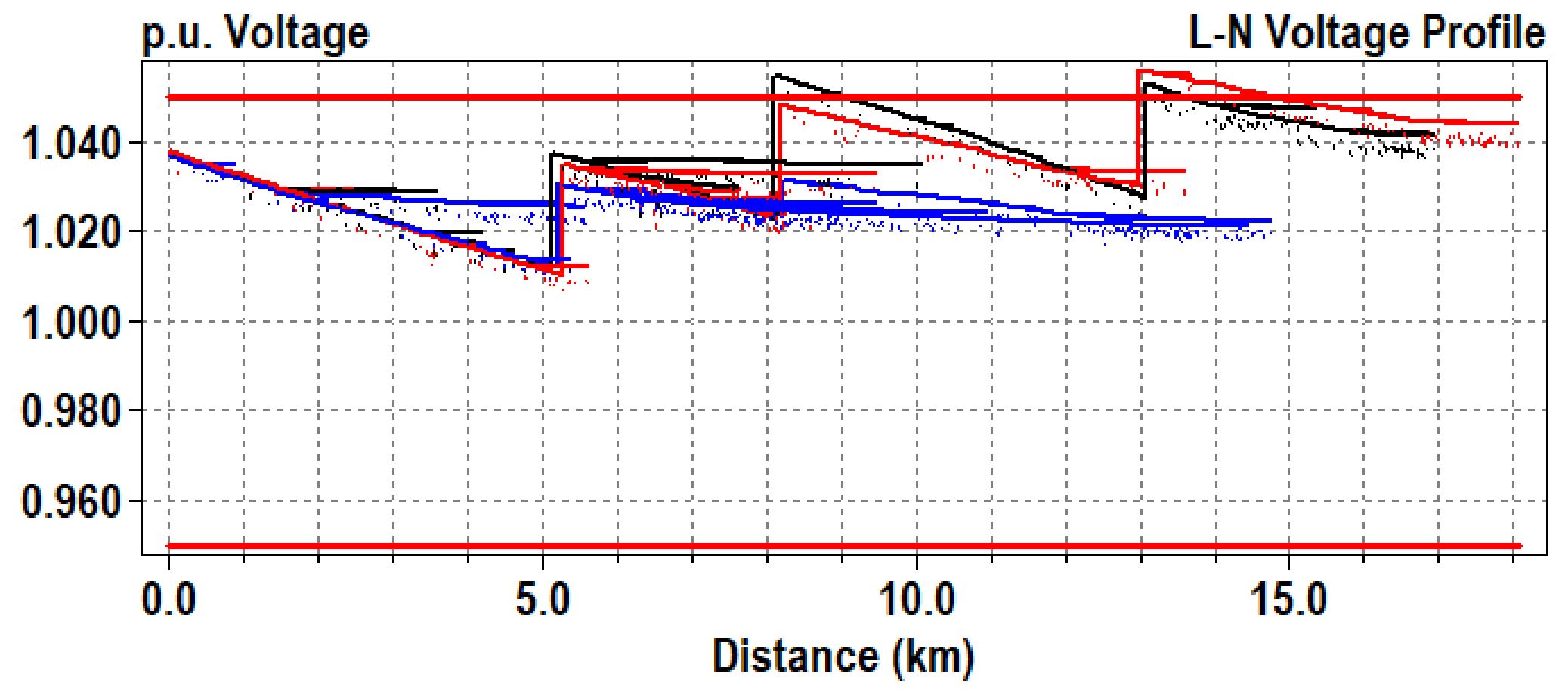}
    \caption{5 AM (Black: Phase A, Red: Phase B, Blue: Phase C) }
  \label{fig: 5AM }
\end{subfigure}
\hfill
\begin{subfigure}[b] {0.48\textwidth}
    \centering
    \includegraphics[width=1\linewidth]{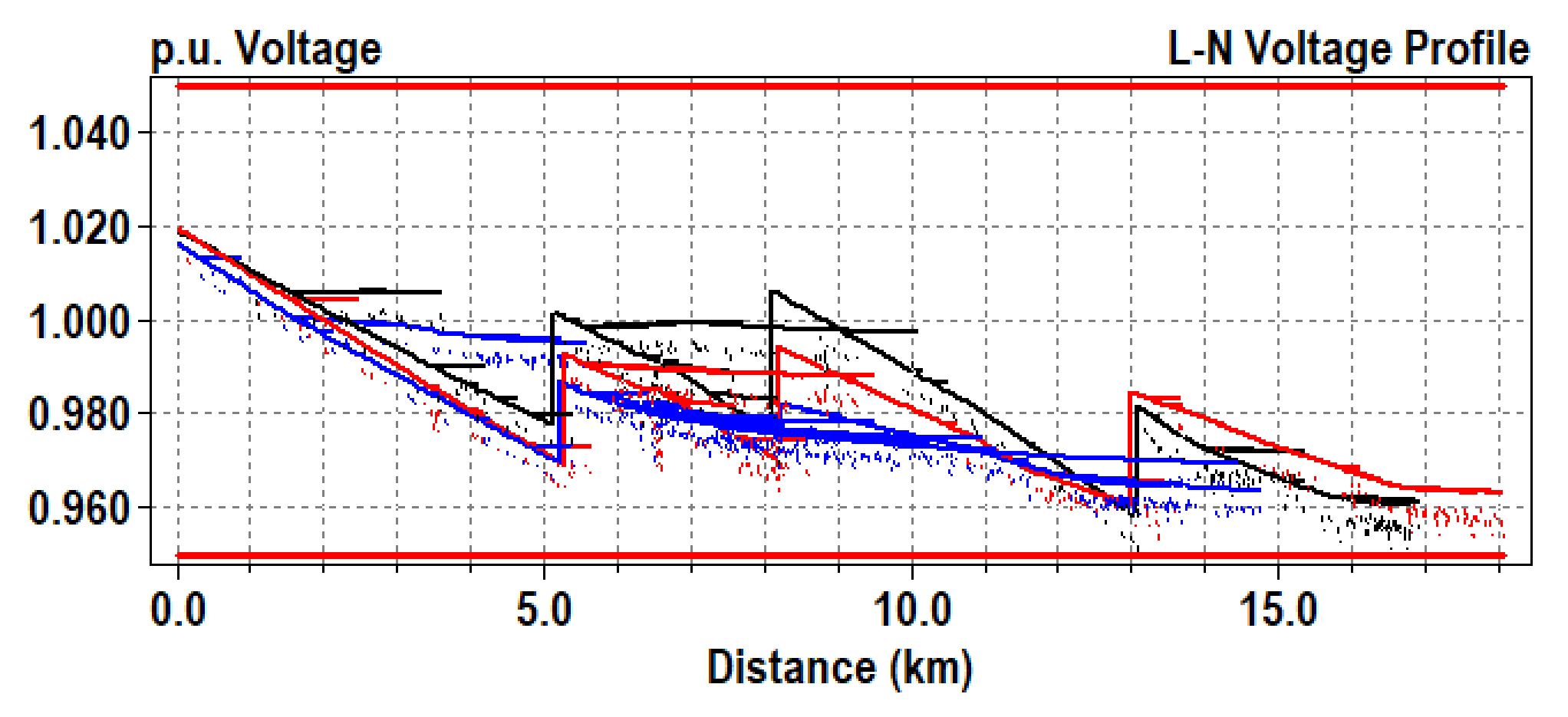}
    \caption{7 PM (Black: Phase A, Red: Phase B, Blue: Phase C)}
  \label{fig: 7PM}
\end{subfigure}    
    \caption{Voltage profile along J1 feeder with selected settings of SVRs/capacitor banks}
  \label{fig: VProf1}
\end{figure}

The outcome of this approach is illustrated in Figs. \ref{fig: VProf1}a and \ref{fig: VProf1}b, where the voltage profiles are shown for the selected SVR and capacitor settings for the month of August. It can be seen that the selected settings do not fully eliminate the violations, but they achieve the best available trade-off with the existing capacitors and SVRs. For example, any single step reduction in any tap settings to address the OV violations at 5 AM results in many more UV violations at 7 PM, and vice-versa. 

The resultant capacitor and SVR settings are fixed for a month and used for all PV HC evaluations. This ensures that the baseline settings are determined without considering the impact of added PVs in the feeder. It should also be noted that the impact of PVs is typically felt more during mid-day hours, while the above-mentioned methodology aims to minimize violations during the periods where PVs have little or no role to play. The resultant capacitor sizes and SVR settings are summarized in Tables \ref{tab:capset} and \ref{tab:regset}, respectively.

\begin{table}
  \centering
  \caption{Capacitor settings summary}
  \setlength{\tabcolsep}{2pt} 
\renewcommand{\arraystretch}{1.5} 
\resizebox{\columnwidth}{!}{%
    \begin{tabular}{|c|c|c|c|c|c|c|c|} \hline
    \textbf{Capacitor ID} & \textbf{Orig. kVAr} & {\textbf{May}} & {\textbf{June}} & {\textbf{July}} & {\textbf{Aug.}} & {\textbf{Sept.}} & {\textbf{Oct.}} \\ \hline
    \textbf{B4909-1} & 900 & 0 & 0 & 900 & 0 & 0 & 0 \\ \hline
    \textbf{B18944} & 1200 & 1200 & 1200 & 0 & 600 & 300 & 600 \\ \hline
    \textbf{B4862-1} & 600 & 0 & 0 & 0 & 0 & 0 & 0 \\ \hline
    \textbf{B4829-1} & 600 & 0 & 0 & 0 & 0 & 0 & 0 \\ \hline
    \textbf{B4877-1} & 600 & 0 & 0 & 600 & 0 & 300 & 600 \\ \hline
    
    \end{tabular}%
    }
  \label{tab:capset}%
\end{table}%

\begin{table}
  \centering
  \caption{SVR settings summary}
  \setlength{\tabcolsep}{2pt} 
\renewcommand{\arraystretch}{1.5} 
\resizebox{\columnwidth}{!}{%
    \begin{tabular}{|c|c|c|c|c|c|c|c|} \hline
    \textbf{Name} & \textbf{Range} & {\textbf{May}} & {\textbf{June}} & {\textbf{July}} & {\textbf{Aug.}} & {\textbf{Sept.}} & {\textbf{Oct.}} \\ \hline
    \textbf{Subxfmr} & .9-1.1 & 1.0125 & 1.0125 & 1.025 & 1.03125 & 1.025 & 1.0125 \\ \hline
    \textbf{B18865} & .9-1.1 & 1.0125 & 1.0125 & 1.0125 & 1.00625 & 1.025 & 1.025 \\ \hline
    \textbf{B19008} & .9-1.1 & 1.03125 & 1.025 & 1.025 & 1.025 & 1.0125 & 1.0125 \\ \hline
    \textbf{B18863} & .9-1.1 & 1.03125 & 1.03125 & 1.04375 & 1.03125 & 1.025 & 1.025 \\ \hline
    \textbf{B19010} & .9-1.1 & 1.0375 & 1.03125 & 1.025 & 1.025 & 1.025 & 1.025 \\ \hline
    \textbf{B4874} & .95-1.05 & 1.025 & 1.025 & 1.025 & 1.025 & 1.01875 & 1.025 \\ \hline
    \textbf{B18864} & .9-1.1 & 1.03125 & 1.03125 & 1.03125 & 1.025 & 1.025 & 1.025 \\ \hline
    \textbf{B4872} & .95-1.05 & 1.025 & 1.025 & 1.025 & 1.025 & 1.01875 & 1.025 \\ \hline
    \textbf{B4868} & .95-1.05 & 1.03125 & 1.03125 & 1.03125 & 1.03125 & 1.025 & 1.03125 \\ \hline
    \end{tabular}%
    }
  \label{tab:regset}%
\end{table}%

\subsection{Baseline Hosting Capacity (HC) Calculations} \label{sub2_5}

Once the SVR settings and capacitor values
are determined, the baseline HC when employing the traditional inverter configurations can be calculated. The two inverter configurations chosen for evaluation were Unity Power Factor (UPF) mode (representing legacy inverters with no control) and volt-VAr (VV) mode (representing the most prevalent existing controlled inverter mode) using the default Category B IEEE 1547-2018 curve (from Table 8 of \cite{8332112}). 

In each case, incremental PV additions were made according to Section \ref{sub2_2}, and power flow was run until the first voltage violation occurred in the feeder. Table \ref{tab:PVadd} summarizes the level of PV additions under different conditions. Given that the peak load is 10.95 MW and the existing PV capacity is 1813.6 kW, these numbers represent HC of 19.3\% and 20.6\%, respectively, for the UPF and VV configurations for the August ``All" set.  
Note that the lowest HC across all configurations 
determines the overall HC of the feeder.  
As expected, the HC numbers for the ``Far" set are lower, and for the ``Near" set are higher.
The violations first tend to occur at hours 10, 11, or 12 across all the months, as those hours represent high (near peak) PV generation and non-peak load.

\begin{table}[htbp]
  \centering
  \caption{PV additions (in kW) prior to voltage violations }
  \renewcommand{\arraystretch}{1.5} 
    \begin{tabular}{| *{7}{c|}} \hline
        \textbf{Case}  & \textbf{May} & {\textbf{June}} & {\textbf{July}} & {\textbf{Aug.}} & {\textbf{Sept.}} & {\textbf{Oct.}} \\ \hline
    \textbf{UPF ``All"} & 660   & 660   & 1100  & 300   & 280   & 560 \\ \hline
    \textbf{UPF ``Far"} & 510   & 520   & 630   & 70    & 210   & 610 \\ \hline
    \textbf{UPF ``Near"} & 1620  & 1260  & 1260  & 480   & 330   & 870 \\ \hline
    \textbf{VV ``All"} & 840   & 850   & 1740  & 440   & 770   & 700 \\ \hline
    \textbf{VV ``Far"} & 630   & 650   & 1380  & 430   & 960   & 790 \\ \hline
    \textbf{VV ``Near"} & 1810  & 1940  & 2800  & 1540  & 1440  & 1200 \\ \hline
    \end{tabular}%
  \label{tab:PVadd}%
\end{table}%

\begin{figure}[!htp]
\centering
  \includegraphics[width=0.48\textwidth]{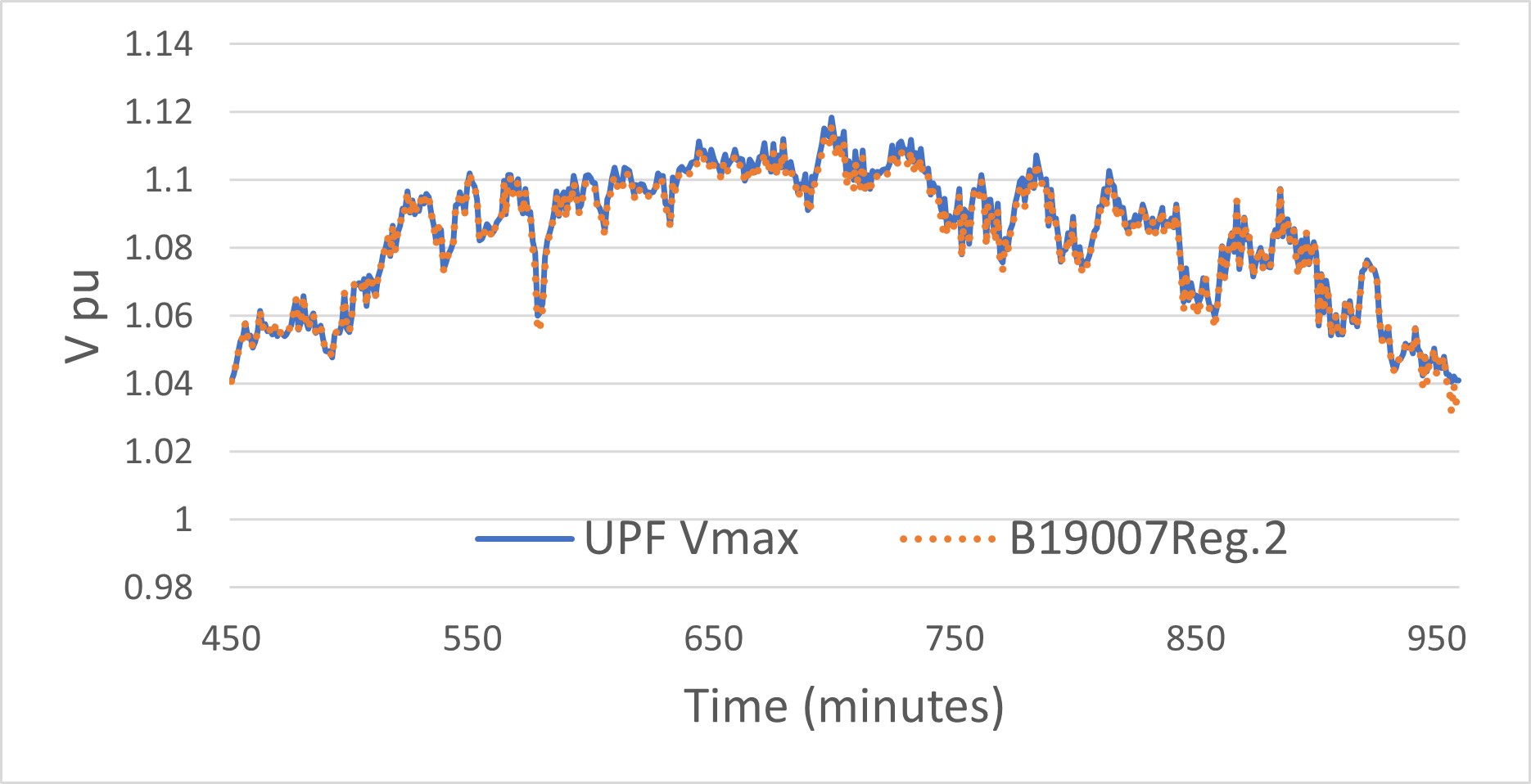}
  \caption{Plot of maximum and B19007reg.2 voltages }
  \label{fig: VProf}
\end{figure}

For this feeder, many of the initial violations were observed at node B190007reg.2, which is the secondary of an SVR on phase B on the far end of the feeder. Fig. \ref{fig: VProf} depicts the profile of this node 
between 7:30 AM and 4 PM, at 1-minute intervals. The figure also shows the maximum voltage at any node on the feeder under UPF conditions for August for a high PV penetration case. The close match between the two profiles indicates that the maximum voltage in the feeder occurs at this node under most conditions. 

\subsection{Discussion on Limitations of VV Control}\label{VoltVAR}
In Table \ref{tab:PVadd},
the lack of significant HC increase when going from UPF to VV configuration is clearly visible. For the J1 feeder, the HC improvement ranges from 1\% (for August, October ``All" and May, June ``Far") to 14\% (for July ``Near"), but the majority of the conditions are lower than 4\%. There are multiple reasons for this finding:

\begin{itemize}
    \item \textbf{SVR settings:} As shown in Fig. \ref{fig: VProf}, the J1 feeder violations mostly occur at 
    the secondary of SVR B19008 that 
    has a step-up ratio of 1.025 in August as per Table \ref{tab:regset}. This setting was identified to be the most effective for the existing J1 feeder in eliminating both UV at high load and OV at light load. However, with PV additions, this step-up causes OV in the secondary. 
    \item \textbf{Lack of system-wide situational awareness:} The VV method may not be able to react to the violation on node B19007reg.2 as it has no awareness of that violation. Similarly, there may be other nodes in the system that may have OV without the VV method being able to react.  
    \item \textbf{Limited Q-intervention under default profile:} The default VV profile in IEEE 1547-2018 activates maximum Q only when PCC voltage is at 1.08 p.u.; this is another limitation of the VV approach.
    The setting can be changed to achieve better results, but it comes with stability concerns as the VV control is not coordinated.  

\end{itemize}

\section{Coordinated Optimization of Inverters for Voltage Support} \label{COIN}
From the analysis conducted in Section \ref{Section2}, it is clear that there is a need for a more effective PV inverter control methodology that is capable of achieving significantly higher HC compared to both UPF and VV-based controls. To implement such a control, a key requirement is the replacement of the PCC information available to each PV inverter 
with global information pertaining to voltage magnitudes of all the nodes of
the system. A coordinated
controller that can process the system-wide voltage information and generate 
prioritized reactive power intervention needs from each of the PV inverters in the system could achieve voltage regulation at much higher PV penetration.
Sections \ref{Sensit_Matrix} to \ref{Algorithm} describe the development of the proposed coordinated control algorithm, while Section \ref{State_Est} describes how ML can be used to attain system-wide voltage information in real-time. Lastly, Section \ref{Algorithm1} presents an alternate approach based on phase-based zoning for comparison with the proposed approach.

\subsection{Application of Sensitivity Matrix (SM)}\label{Sensit_Matrix}
The core of the proposed control algorithm is 
the voltage-reactive power (VQ) SM. 
An SM provides an indication of the voltage response capability of each bus in the network to a unit active or reactive power perturbation at selected nodes \cite{Zhu}. For the proposed control, only the SM for Q is employed, as the goal is to improve voltage\textit{ without any change} in the active power. The SM is a matrix with $M$ columns corresponding to the number of PV inverters and $N$ rows corresponding to the number of nodes in the system, as shown in \eqref{s_mat}. 

\begin{equation}\label{s_mat}
SM = 
\begin{bmatrix}
    sm_{1,1} & sm_{1,2} & \dots & sm_{1,M} \\
    sm_{2,1} & sm_{2,2} & \dots & sm_{2,M} \\
    \vdots & \vdots & \ddots & \vdots \\
    sm_{N,1} & sm_{N,2} & \dots & sm_{N,M} 
\end{bmatrix}
\end{equation}

The SM is constructed by computing each element using \eqref{s_comp}, where $\Delta Q_j$ is the reactive power perturbation applied to PV inverter $j$, $v_i^{0}$ is the voltage at node $i$ prior to the perturbation and $v_{i,j}^Q$ is the voltage at the same node after the perturbation. 

\begin{equation}\label{s_comp}
    sm_{i,j} = \frac{v_{i,j}^Q - v_i^{0}}{\Delta Q_j}
\end{equation}

Generation of SM leads to a linear set of equations that relate the voltage deviations achievable in the distribution network to the individual inverter's Q injection/absorption. This set of equations is utilized in the control algorithm to calculate the optimum Q contribution from each inverter to regulate the node voltages. However, two additional factors must be considered for this methodology to be applied to high HC cases in a robust manner:

\begin{itemize}
    \item The SM contains entries of both polarities.
    In other words, a reactive power perturbation at a single PV can have an additive effect on the voltage of some nodes and a subtractive effect on other node voltages. 
    \item The SM values are not static. While it is computationally useful to assume fixed sensitivity values, the reality is that the sensitivity is dependent on the actual operating conditions at each node, including both active and reactive power values, as well as the voltage levels. 
\end{itemize}

Both these factors have not been sufficiently addressed in prior voltage control algorithms \cite{Lusis, 8973721}, leading to algorithms that have limited robustness/HC ability.
In this study, both these impacts are analyzed and addressed 
to provide consistently better results across a large number of scenarios. 

\subsection{Cross-phase Sensitivity in Unbalanced Networks }\label{Cross_Phase}
The bipolar nature of the SM entries is explainable by considering that the distribution networks are not inherently balanced. The unbalanced loading conditions lead to significant non-zero currents in the neutral conductors. If reactive power perturbation is made on a single phase node, the resultant neutral current flow will have different effects on the voltages in each of the other phase nodes close to the perturbation. 
For an absorption of $\Delta Q_A$ on Phase A, the currents $\Delta I_A$ and $\Delta I_N$ are orthogonal to $V_A$. With mainly reactive impedance in each path, the resultant $\Delta V_A$ has the opposite polarity of $V_A$, as desired. 
However, the same $\Delta Q_A$ and resultant  $\Delta I_N$ can have a different impact on $V_B$ and $V_C$. This effect is dominated by the neutral path current, but the mutual coupling between the conductors also plays a role. When the phasor additions are done for Phases B and C, $\Delta V_B$ and $\Delta V_C$ exhibit an increase in magnitude as a result of $\Delta Q_A$ absorption on Phase A in many instances. In other instances, one of the two phases shows an increase, while the other phase shows a slight decrease. 

The consequence of this phenomenon 
is that any reactive power intervention in one phase to help address the voltage violation of that phase can have an opposite effect in a nearby node with a different phase leading to non-mitigation of the voltage violation in the other phase and/or creation of a new voltage violation in the other phase. Under high PV penetration, since many of the voltage violations can be geographically close to each other, the impact of this phenomenon can be profound.
Finally, note that this cross-phase sensitivity has not been addressed in any prior literature on HC analysis. 

\subsection{Iterative SM Refinement}\label{Iterative}
The elements of the VQ SM are calculated by providing unit reactive power (Q) perturbation to each PV node and evaluating the voltage change using power flow simulations. However, the results of the power flow are highly dependent on the operating points of the network. Assuming that the starting point is the UPF operation of all PVs, all the inverters are at zero Q. The creation of SM from these conditions may be very different from the final conditions, where many PVs may have substantial non-zero Q values. Moreover, if the initial voltages are far from 1.0 p.u. (as may happen during certain hours under high PV penetrations and no Q intervention), the generated SM is also going to be inaccurate. 
Prior literature on SM did not consider these factors.
The proposed iterative solution recomputes the SM using Q values of the previous iteration prior to running the optimization for each iteration, leading to optimal Q intervention to address all voltage violations. 



\subsection{Proposed Optimization Algorithm}\label{Algorithm}
Based on the above-mentioned considerations, we propose an
iterative linear optimization formulation as shown in Algorithm \ref{Alg_1}. Steps 1 to 9 of this algorithm help establish the base conditions for evaluation and initial results, which may or may not indicate the need for Q intervention. If an intervention is required based on the test conditions in Step 10, the SM will be calculated by solving \eqref{s_comp} using the current values of the \textbf{Q} vector. Next, the linear optimization given in \eqref{opt_eq} will be performed with the objective of minimizing total Q intervention from PV inverters. 



\begin{equation} \label{opt_eq}
    \begin{aligned}
    \text{Minimize} & \quad\sum_j (Q_j + \Delta Q_j) \\
     \text{subject to} \: & { -Q_{jmax} \le (Q_j + \Delta Q_j) \le Q_{jmax}} \quad \forall j\\
            & {Vpu_{min} \le (V_i + \Delta V_i) \le Vpu_{max}} \quad \forall i\\
    & {\Delta V_i = \sum_j sm_{i,j} \times \Delta Q_j} \quad \forall i,j
    \end{aligned}
\end{equation}

In \eqref{opt_eq}, $Q_j$ is the $j^{th}$ element of current $\mathbf{Q}$ vector representing the $M$ PV inverters' reactive power contribution; $\Delta Q_j$ is the $j^{th}$ element of the  $\mathbf{\Delta Q}$ vector calculated by the optimization algorithm; $-Q_{jmax}$ and $Q_{jmax}$ are the minimum and maximum values of reactive power available from an inverter; $V_i$ is the $i^{th}$ element of the $\mathbf{V}$ vector that represents the $N$ nodes' voltage magnitudes (in p.u.); $\Delta V_i$ is the $i^{th}$ element of the $\mathbf{\Delta V}$ vector representing the permissible or required change in the $V_i$ value to keep it within regulation with some guard-bands; $Vpu_{min}$ and $Vpu_{max}$ are the guard-banded voltage limits for the network.
The minimization formulation given in \eqref{opt_eq} helps accomplish following goals:
(a) minimize total stress on inverters; (b) contain additional losses caused by Q circulation; and (c) minimize communication burden by limiting the number of PVs that must be controlled.

\begin{algorithm}[t]
\begin{algorithmic}[1]
\caption{Iterative Optimization Algorithm} \label{Alg_1}
\State Load network information into a distribution system solver
\State Set SVR, OLTC, and/or capacitor bank settings
\State Add PV systems 
\State Apply appropriate PV and load profiles
\State Select evaluation condition (time instance)
\State Determine base Q values: 
\begin{equation*}
    \mathbf{Q^0} = [Q_0, Q_1, \dots Q_j, \dots, Q_M] 
\end{equation*}

\State Run power flow with base conditions and calculate base voltage magnitudes: $\mathbf{V^0} = [V_0, V_1, \dots V_i, \dots, V_N] $
\State $ \mathbf{V} \gets \mathbf{V^0}$
\State $ \mathbf{Q} \gets \mathbf{Q^0}$
\While{($\max(\mathbf{V}) > V_{thmax}$) or ($\min(\mathbf{V}) < V_{thmin}$)} 
    \State Calculate Sensitivity Matrix, $\mathrm{SM}$, with $\mathbf{Q}$ values
    \State Solve \eqref{opt_eq} to determine $\mathbf{Q^n} = \mathbf{Q}+ \mathbf{\Delta Q}$

    \State Apply $\mathbf{Q^n}$ values to PV and run power flow to get $\mathbf{V^n}$
    \State $ \mathbf{V} \gets \mathbf{V^n} $
    \State $ \mathbf{Q} \gets \mathbf{Q^n} $
\EndWhile
\end{algorithmic}
\end{algorithm}

The first constraint ensures that the reactive power values for each PV are consistent without requiring APC under any conditions. In other words, the values of $Q_{jmax}$ are determined by  \eqref{eq:Qmax}, where $S_j$ is the apparent power (kVA) rating of the PV and $P_{jmax}$ is the $P_{mpp}$ of the $j^{th}$ PV. 
\begin{equation} \label{eq:Qmax}
    Q_{jmax} = \sqrt{{S_j}^2 - {P_{jmax}^2}}
\end{equation}
For all the added PVs, the value of $Q_{jmax}$ was calculated to be 4.8 kVAr for 10 kW  $P_{jmax}$ rating \cite{8332112}. If additional voltage support is needed, this constraint can be made into a non-linear constraint by incorporating the actual value of $P_j$ at a given time instant in \eqref{eq:Qmax} in place of $P_{jmax}$.

The second constraint sets the desired targets for the optimization problem to meet the voltage regulation target. The targets in this constraint are tighter than the test condition in Step 10 to cover for any differences in the power flow results from idealized calculations based on SM as well as errors in the estimates of the system-wide voltage information.

The final constraint recognizes that the voltage correction at each node is the sum of the product of each Q perturbation in the network and the sensitivity of the node to that perturbation. In an unbalanced distribution network, this formulation must include both positive and negative values of $sm_{i,j}$ to ensure accurate results. 
The positive values arise from the cross-phase impacts, while the negative values arise from the in-phase impacts.

\subsection{Achieving System-Wide Situational Awareness with Limited Sensor Coverage}
\label{State_Est}
The sensors present in a modern distribution system are primarily located at the root node (SCADA and PMUs/$\mathrm{\mu}$PMUs at feeder-head) or at the leaf nodes (smart meters and solar meters at residences). As such, bulk of the feeder is invisible to power utilities from a monitoring and controls perspective.
This unobservability due to lack of sensor coverage also impacts the performance of conventional approaches for distribution system state estimation.

To attain system-wide situational awareness with limited sensor coverage, the ML-based time-synchronized state estimator developed in \cite{azimian2022state} can be used. 
It employed Bayesian inference to indirectly learn a mapping relation between fast timescale feeder-head PMU/$\mathrm{\mu}$PMU measurements and the state (voltage magnitude and angle) of every node of the system.
This state estimator had a mean absolute percentage error of $<0.03\%$ in predicting nodal voltage magnitudes within milliseconds for a complex real-world distribution system located in a metropolitan city of the U.S. Southwest.
More details about this ML-based state estimator can be found in \cite{azimian2023book}.

Lastly, note that for the simulations done in this paper, the voltage values before and after applying the proposed control algorithm were obtained using OpenDSS \cite{OpenDSS}, which is a distribution system solver. 
That is, the outputs of OpenDSS were a proxy for the outputs of the ML-based state estimator. 
During an actual implementation in the real world where we do not have full observability, we will first create the ML-based state estimator to provide the input as well as verify the output of the proposed control algorithm. 


\subsection{Algorithm Modification for Independent Phase Operation}\label{Algorithm1}
A significant set of prior research in the field of voltage regulation has used different zoning methodologies to achieve faster-decentralized control \cite{Zon_1, Zon_2,Zon_3}. Creation of the
zones by phase (Phases A, B, and C) is the most popular approach as it represents the results of common clustering algorithms applied to the SM. Hence, Algorithm \ref{Alg_1} was modified by separating the PVs and loads by phase, creating sensitivity matrices for each phase, and running the optimization for each phase independently; this modified algorithm is henceforth referred to as the ``Zoned" Algorithm. 

\section{Analysis of Results} 
\label{Res_section}
Algorithm \ref{Alg_1} was applied to the J1 feeder along with the PV addition methodology described in Section \ref{sub2_2}, Pecan Street profiles mentioned in Section \ref{sub2_3}, and capacitor and SVR settings determined in Section \ref{sub2_4}. The number of PVs was incrementally increased till the point where the algorithm was unable to find voltage mitigation solutions across different use cases. 
The power flows were performed in OpenDSS \cite{OpenDSS}.
$\mathbf{Q^0}$ was set to
zero to represent the worst-case scenario where the control algorithm is engaged for the first time without any pre-existing Q intervention. This allowed the algorithm to be tested under the harshest conditions where it is required to address a very large number of violations ($>70$\% of the nodes) and high voltage excursions ($ \max (\mathbf{V^0}) > 1.1 \: \mathrm{p.u.}$). 

\subsection{Hosting Capacity (HC) Improvement}
The ``All" set was considered first as it represents the most likely approach for PV addition.
Moreover, we focused on the ``All" set for August as it exhibited the lowest base case HC.
The total PV penetration without any violations improved to 6.41 MW (while limiting the number of optimization iterations to three) for this use case as opposed to 2.25 MW for VV and 2.11 MW for UPF - representing a 285\% increase in PV generation. The improvement in HC is also about  3x, as shown in the last row of Table \ref{tab:HC_improv}. 
The Zoned Algorithm resulted in shorter computation times while also providing Q intervention solutions for a high HC value. However, when applied to the OpenDSS power flow, these Q intervention values resulted in many voltage violations
due to the disregarding of the cross-phase effects. 
The actual HC results for this approach, determined by the highest PV levels under which the OpenDSS power flow 
gave zero violations, were only slightly better than the VV or UPF results (see ``Zoned" column of Table \ref{tab:HC_improv}),
and significantly worse than those achieved with Algorithm \ref{Alg_1}
(see column ``Optimized" of Table \ref{tab:HC_improv}).

\begin{table}[htbp]
  \centering
  \caption{HC improvement with proposed algorithm}
  \setlength{\tabcolsep}{2pt} 
  \renewcommand{\arraystretch}{1.5}
    \begin{tabular}{|l|c|c|c|c|c|} \hline
          & \multicolumn{4}{c|}{\textbf{Hosting Capacity (HC)}} & \textbf{Improvement} \\ \hline
          & {UPF} & {VV} & {Zoned} & {Optimized} & {$\Delta$ to VV} \\ \hline
    kW    & 2113.6 & 2253.6 & 2983.6 & 6413.6 & 4160 \\ \hline
    \% of Peak Load & 19.3\% & 20.6\% & 27.2\% & 58.6\% & 38.0\% \\ \hline
    \end{tabular}%
  \label{tab:HC_improv}%
\end{table}%

\subsection{Sensitivity Variation}

\begin{figure}
\centering
    \includegraphics[width=1\linewidth]{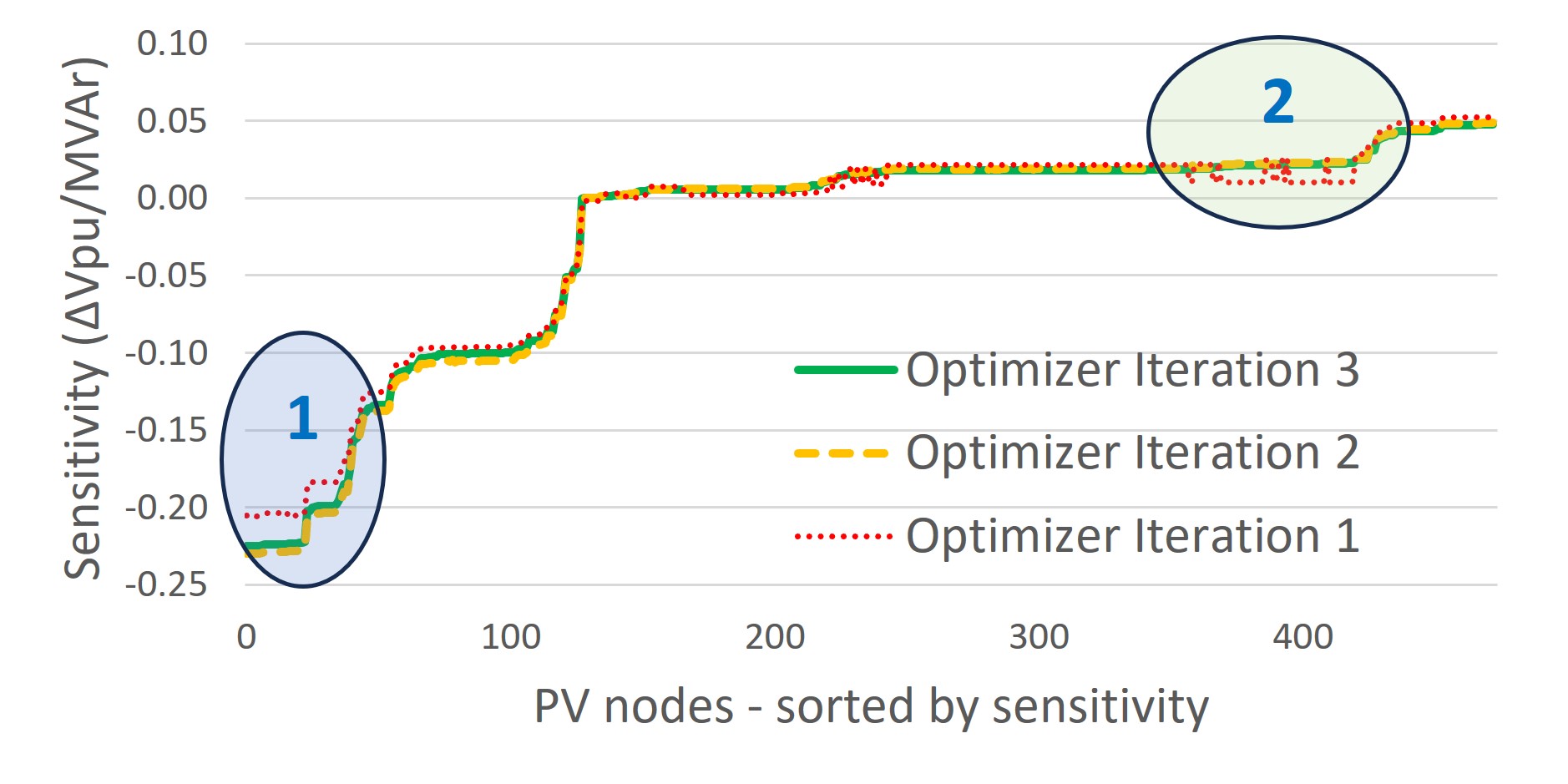}
    \caption{Sensitivity of B19007reg.2 to all PVs in the network. The ovals indicate the differences in the sensitivities of the first iteration (for which Q is zero) and the second and third iterations (for which Q is non-zero)}
  \label{fig: Sensit}
\end{figure}

The effects of cross-phase sensitivity and the iterative SM refinement are clearly visible in Fig. \ref{fig: Sensit}, which depicts sensitivity for node B19007reg.2. 
Note that this node had the greatest need for voltage correction (as observed in Fig. \ref{fig: VProf}). The sensitivity values are plotted for three optimization iterations, and it shows that the first iteration values differ significantly from the second and third iterations. It can also be seen that the negative sensitivity values are higher in magnitude, indicating the in-phase effects. The positive values indicate the cross-phase effects, and even though the values are smaller, their cumulative impact cannot be ignored.

\subsection{Hourly Results and Analysis}
\begin{figure}[b]
\centering
    \includegraphics[width=1\linewidth]{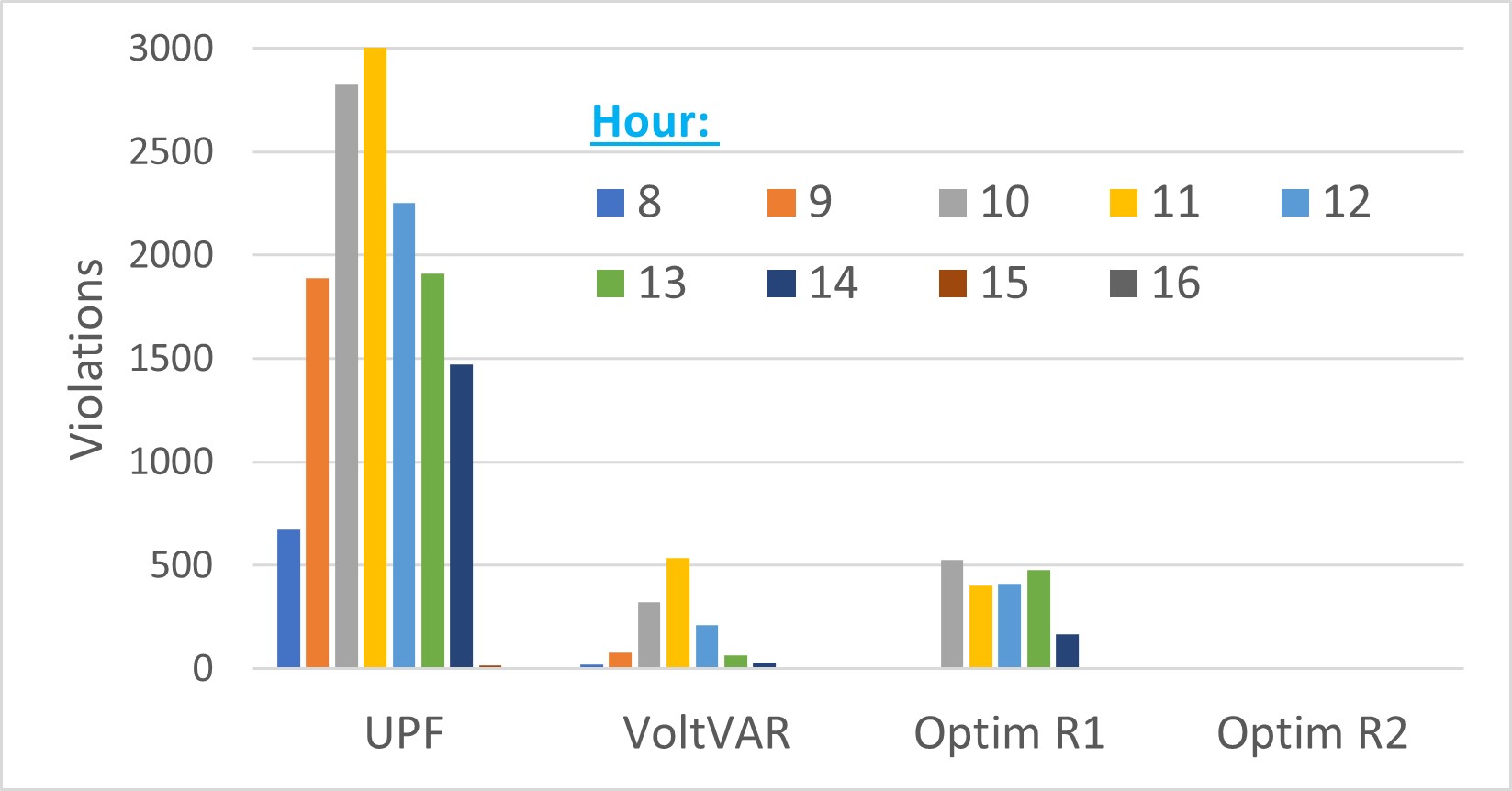}
    \caption{Aug. voltage violation count for different conditions}
  \label{fig: Viol}
\end{figure}

\begin{figure}
\centering
    \includegraphics[width=1\linewidth]{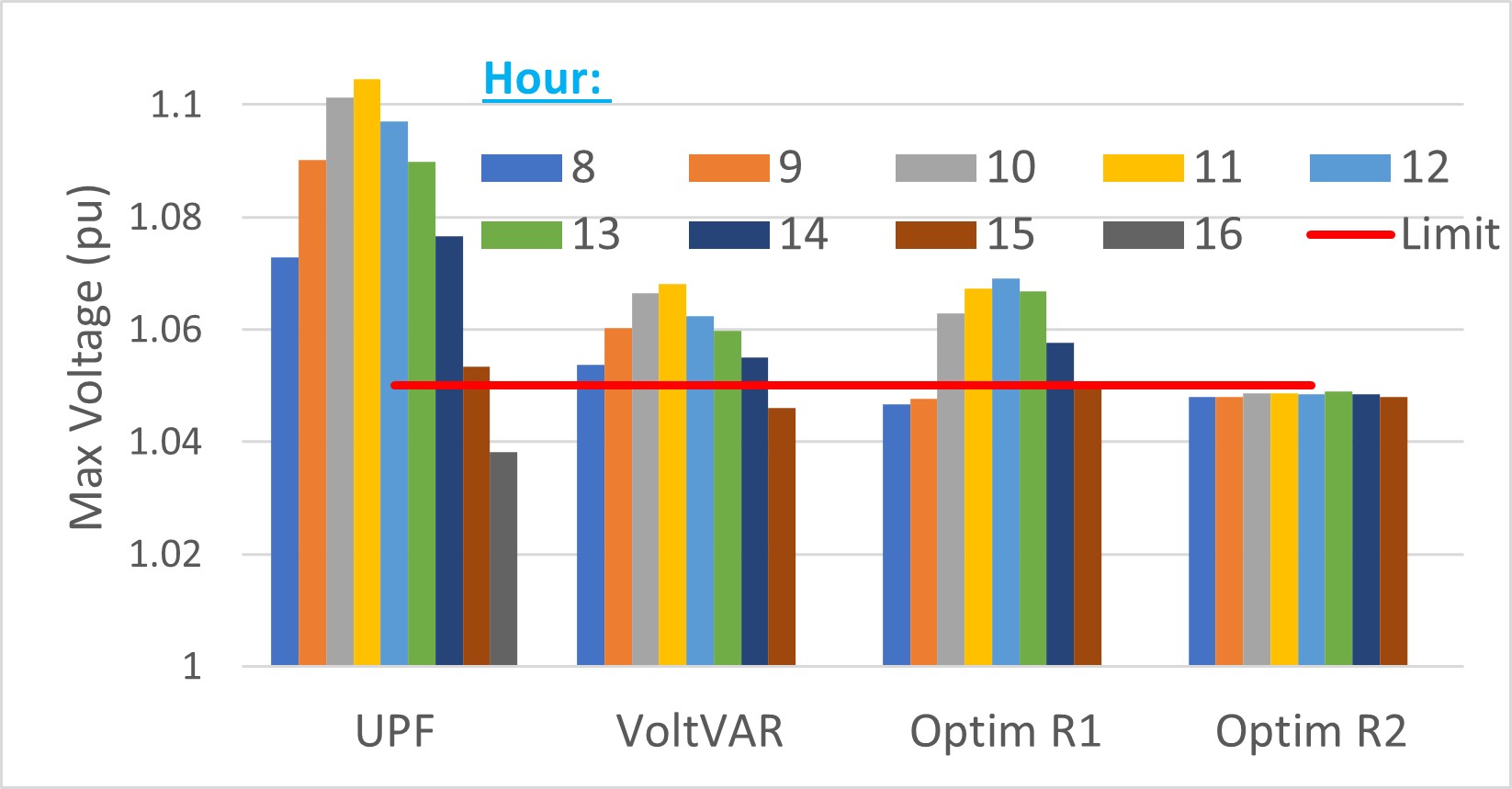}
    \caption{Aug. maximum p.u. voltage for different conditions}
  \label{fig: MaxPU}
\end{figure}

\begin{figure}
\centering
    \includegraphics[width=1\linewidth]{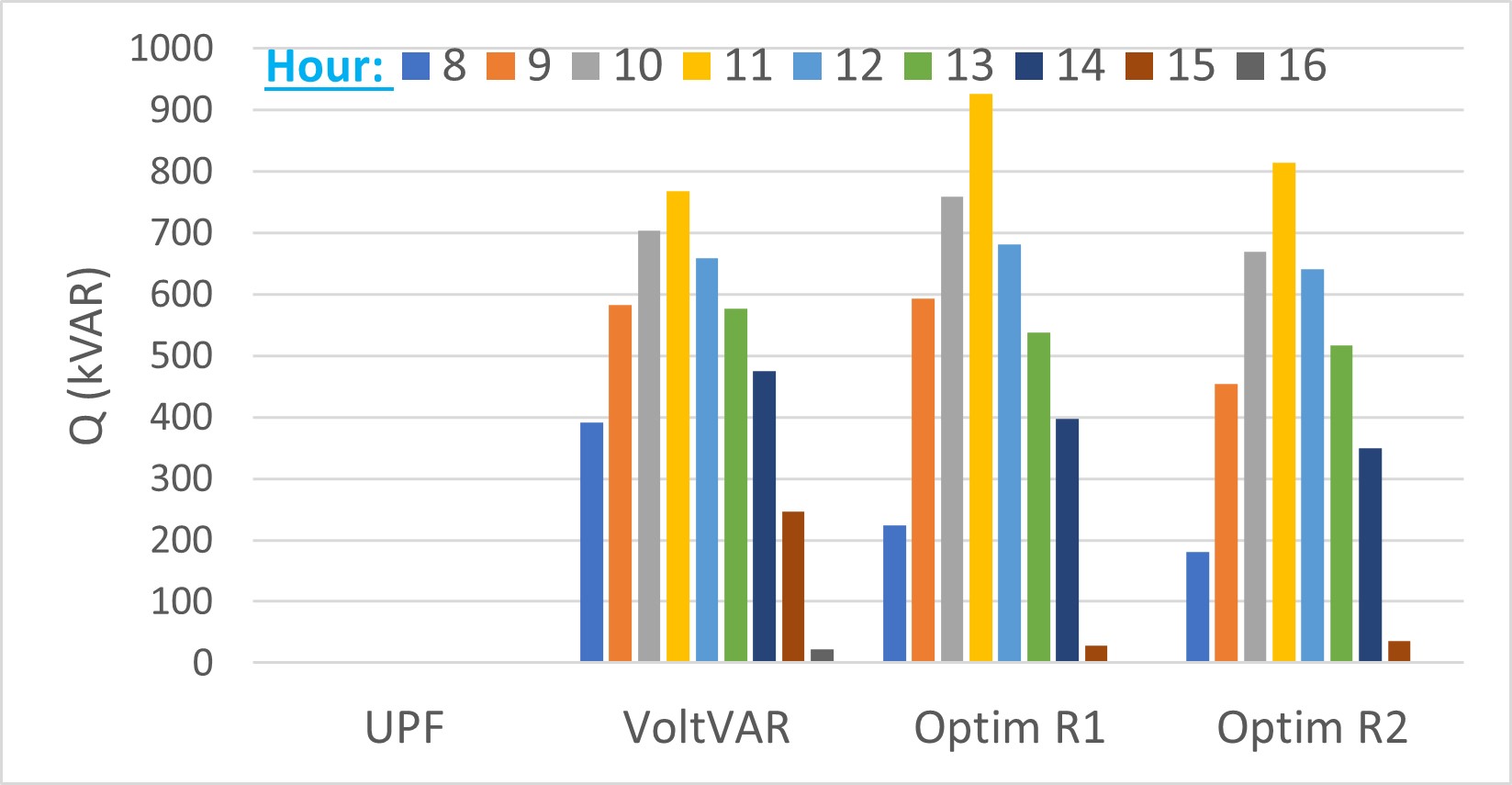}
    \caption{Aug. Q support for different conditions}
  \label{fig: Q_tot}
\end{figure}

Figs. \ref{fig: Viol}-\ref{fig: Q_tot} show the number of OV violations, maximum voltage, and required total Q, respectively, in the J1 feeder, with the reported high PV penetration (6.41 MW) across different hours and under different control settings (Optim R1 and Optim R2 refer to the first and second iterations of the optimization algorithm). It can be seen that the 3042 violations under UPF at 11 AM are reduced to 534 by employing VV control, along with the reduction in maximum voltage from 1.1045 p.u. to 1.0681 p.u. However, the optimization algorithm is able to eliminate all the violations in the second iteration and restrict the maximum voltage to 1.0485 p.u. The total Q support at 11 AM is 814 kVAr for Optim R2 compared to 768 kVAr for VV control. 
While both numbers are comparable, the superior results for the optimization algorithm are attributable to its ability to maximize Q support from the inverters that contribute the most to the required voltage correction. This pattern is repeated at different hours but with reduced levels of Q for each data point compared to the peaks at 11 AM.

Furthermore, these figures show that when going from Optim R1 to Optim R2, not only do the number of violations and the maximum voltage values come down as desired, but also they are accompanied by a reduction in the total Q value. 
This is a seemingly counter-intuitive result since higher Q absorption is required to reduce voltages.
The explanation for this result is provided by considering the sensitivity values and the cross-phase voltage effects of Q absorption. Referring to Fig. \ref{fig: Sensit}, the sensitivity values for iteration 2 and 3 are closer to each other and they represent higher accuracy (because the Q values are non-zero). Comparatively, iteration 1 values are lower in magnitude for the same phase (oval 1) - indicating that to achieve the same $\Delta V$, higher Q will be calculated by the optimization algorithm for the in-phase PVs. However, as shown in oval 2, the cross-phase PV sensitivity values for iteration 1 are also relatively lower in magnitude for many PVs - indicating that the optimization program will underestimate the subtractive effect of Q intervention at those PVs on the given node - allowing the Q in other phases to also go high. The end result is that the first iteration will often provide higher Q without the ability to completely mitigate voltage violations in the network.     

This result also points to the practical complexity of achieving the required voltage mitigation because simply increasing the absorbed Q in an uncoordinated manner is very likely to result in failure to eliminate voltage violations. In fact, this phenomenon was clearly observed by gradually increasing the Q values in one phase - it led to a reduction in maximum voltage in that phase, but beyond a certain point, it started causing more violations in the other phases. 

\subsection{Results for Other Months}
While the application of the methodology to the worst case period (August) was validated across all hours of PV generation as shown in Figs. \ref{fig: Viol}-\ref{fig: Q_tot}, it was still imperative to check the validity of the HC across all months and hours. This was accomplished by employing Algorithm \ref{Alg_1}
for all the months but with cognizance of the capacitor and SVR settings as listed in Tables \ref{tab:capset} and \ref{tab:regset}. It was noticed that as the months vary, the UPF worst-case conditions also vary in terms of hour of occurrence, number of violations, dominant phases for the violations, and maximum voltage. The UPF violations spread is depicted in Table \ref{tab:UPF_mon}, which also includes information about the violations observed in the secondary and at PV locations under the given conditions. Since about 50\% of the violations occur on the primary side, the VV control is not always able to mitigate those violations as it does not sense any impact on the PV nodes. The variations in Table \ref{tab:UPF_mon} are indicative of the inherent complexity in achieving effective control of complex distribution networks under high PV penetrations in the absence of system-wide situational awareness. 

\begin{table}[htbp]
  \centering
   \setlength{\tabcolsep}{4pt} 
  \caption{Summary of UPF worst-case violations }
   \renewcommand{\arraystretch}{1.5}
    \begin{tabular}{|c|c|c|c|c|c|c|c|} \hline
     & & \multicolumn{3}{c|}{\textbf{Phase}} & & & \\ \hline
    Month & {Hour} & {A} & { B} & { C} & {Secondary} & {PV} & {Max V} \\
    \hline
    May   & 12    & 361   & 506   & 430   & 697   & 171   & 1.0907 \\ \hline
    June  & 10    & 363   & 667   & 448   & 775   & 203   & 1.0911 \\ \hline
    July  & 12    & 391   & 838   & 488   & 904   & 248   & 1.0908 \\ \hline
    Aug.  & 11    & 512   & 1013  & 1517  & 1642  & 367   & 1.1045 \\ \hline
    Sept. & 11    & 376   & 1006  & 541   & 1050  & 285   & 1.0936 \\ \hline
    Oct.  & 11    & 361   & 503   & 416   & 679   & 155   & 1.0764 \\ \hline
    \end{tabular}%
  \label{tab:UPF_mon}%
\end{table}%

\begin{figure}[b] 
\centering
    \includegraphics[width=1\linewidth]{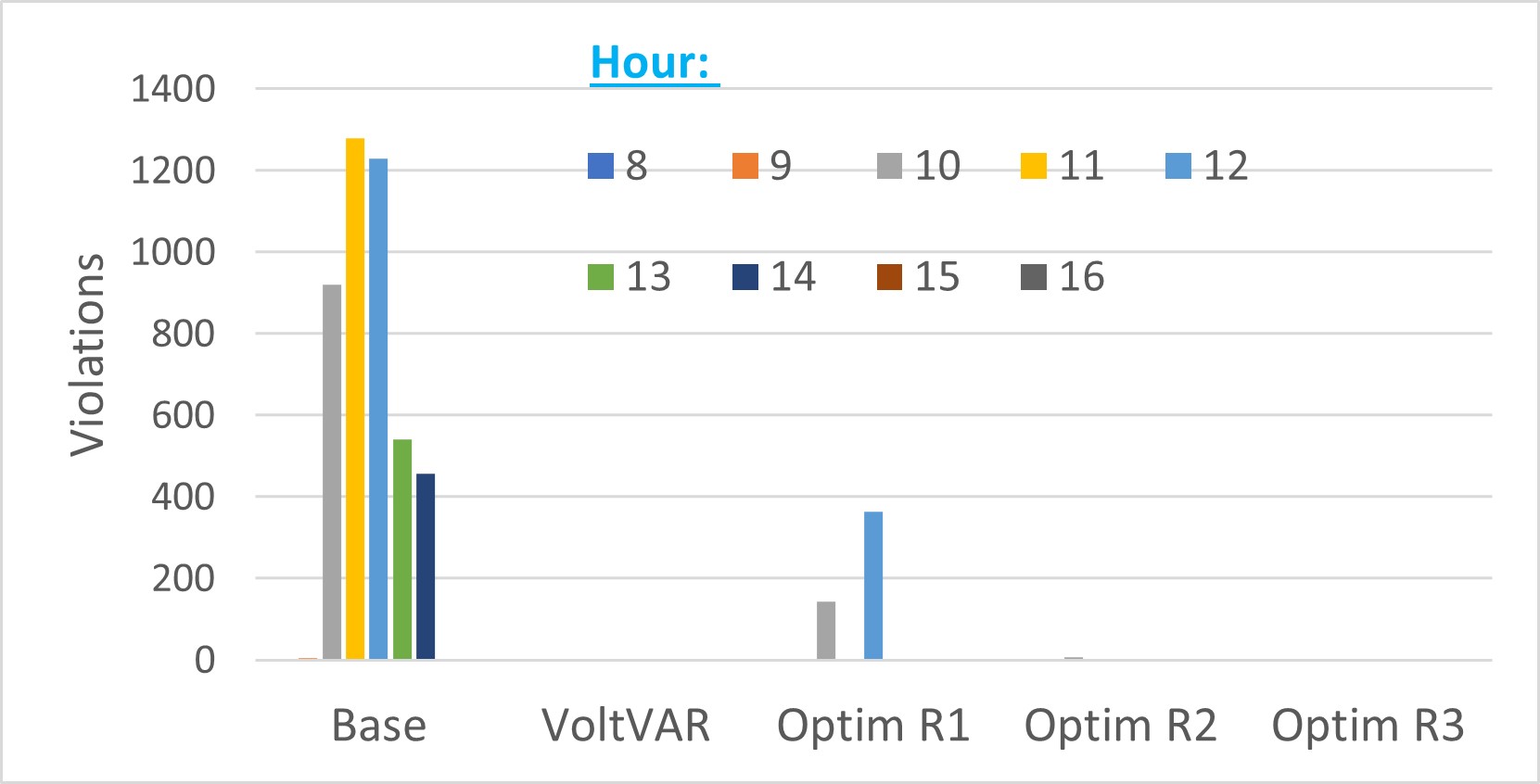}
    \caption{Oct. voltage violation count for different conditions}
  \label{fig: Viol_Oct}
\end{figure}

\begin{figure} 
\centering
    \includegraphics[width=1\linewidth]{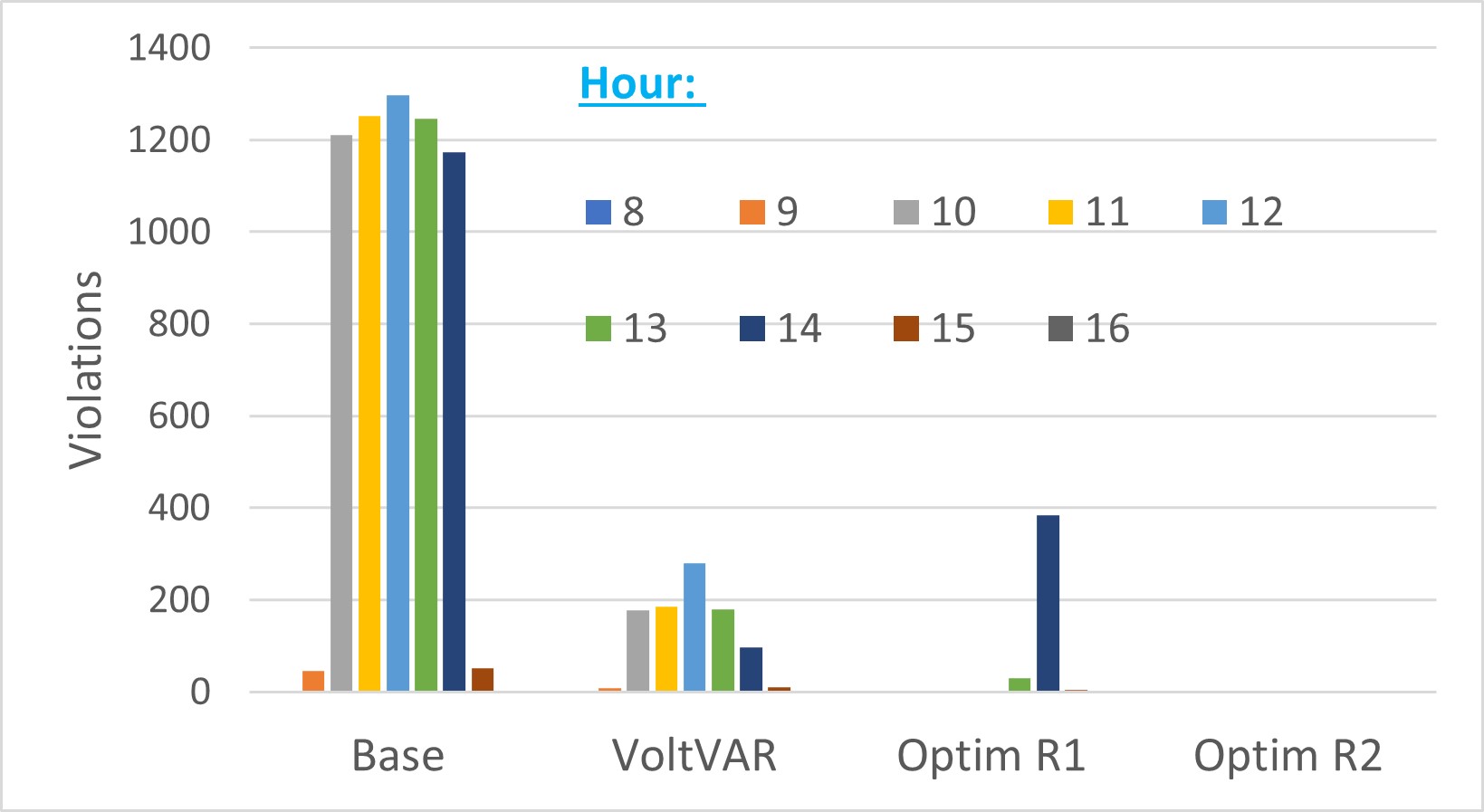}
    \caption{May voltage violation count for different conditions}
  \label{fig: Viol_May}
\end{figure}

The results across the months for both the VV control and the proposed optimization-based control track the UPF results in terms of required Q support. While the proposed optimization algorithm is able to remove voltage violations for all hours in all months, the same is not true for the VV control. For example, in May and October, with less than half the UPF violations compared to August, the required Q values were also reduced by half. However, since the maximum voltage for October was much lower, even VV was able to mitigate all the violations for all the hours (see Fig. \ref{fig: Viol_Oct}), but it failed to do so for May (see Fig. \ref{fig: Viol_May}) for seven of the nine hours. 

\subsection{Comparison with Prior J1 Feeder Results}
One of the reasons for choosing the J1 feeder as the test system for validating the proposed methodology was the availability of detailed HC results for this feeder from \cite{osti_1432760}.
Before providing a results comparison, the similarities and differences between our work and \cite{osti_1432760} are summarized:

\begin{itemize}
    \item \textbf{PV addition methodology:} Both works follow the methodology of ``Near", ``Far" and ``All" additions. Our work standardizes the step addition of PV capacity at 10 kW, while \cite{osti_1432760} used a more stochastic approach. 
    \item\textbf{VV curve:} We used the default settings in IEEE 1547-2018 as opposed to more aggressive settings used by \cite{osti_1432760}. 
    \item \textbf{Inverter Q/PF:} We used variable Q (capped by maximum Q equivalent of 0.9 PF). Ref. \cite{osti_1432760} set a fixed PF of 0.95 absorbing or a VV curve. 
    \item \textbf{Load-PV profiles:} We used a wide range of profiles in a time-varying format available from a real-world database. Ref. \cite{osti_1432760} used a combination of minimum load (undefined) and maximum PV generation.
    \item \textbf{SVR and capacitor settings:} We fixed the settings for a given month and did not allow them to participate in voltage regulation in response to the consequences of added PV generation; \cite{osti_1432760} did the same. 
    \item \textbf{HC basis:} We included the 5 MW aggregated load in all of our simulations and computations. However, \cite{osti_1432760} ignored it in the reported HC calculations (resulting in higher \% values)
\end{itemize}

Despite some key differences, it is still instructive to compare the results of the two studies. For example, \cite{osti_1432760} reported an HC of 1.9 MW with fixed PF control and/or VV control, which is close to the reported value of 2.2 MW in this work. 
Even with allowing the SVR and capacitor settings to change, the prior work could only achieve an HC value of  3.94 MW (Table 7 of \cite{osti_1432760}), while the proposed approach is able to go up to 6.41 MW without any SVR or capacitor adjustments. 

While the results shown so far have focused on the ``All" set, it is also instructive to take a look at the ``Far" set, where the PV additions are concentrated on the far end of the feeder. Ref. \cite{osti_1432760} reported that the techniques employed in that research have very marginal effectiveness in improving the HC, as compared to the ``All" set (361 kW increase of HC). This result is attributable to the inability of the traditional approaches, such as SVR/OLTC/capacitor setting changes, to have a significant impact on the nodes at the edge of the network. In the proposed research, the HC under the ``Far" set is extended from 1.88 MW under the UPF case to 4.91 MW. While this increase is not as large as the ``All" set, it is still significant and demonstrates the effectiveness of using the system-wide situational awareness in conjunction with a robust control algorithm. 

Lastly, additional validation was also carried out based on minutely
load and PV profiles from Pecan Street. These results were consistent with the results shown for hourly data but provided more granularity and a much larger set of use cases.

\section{Conclusion and Future Work} \label{Concl_section}
Traditional inverter control mechanisms have limited effectiveness in mitigating voltage violations in distribution networks under high levels of DER penetration. Using SVRs, OLTCs, and capacitors to participate in voltage regulation in PV-rich systems has its own downsides (e.g., increased loss-of-life due to frequent usage). A novel coordinated control methodology was implemented in this paper that overcame these limitations and \textit{tripled} the HC in the EPRI J1 Feeder under a broad set of realistic test conditions. 
The results are attributable to three core improvements: 
\begin{itemize}
    \item Use of system-wide voltage information 
    \item Recognition of cross-phase sensitivity effects and their incorporation in the optimization algorithm
    \item Iterative refinement of sensitivity matrix to have more accurate results from the optimization algorithm
\end{itemize}



The successful implementation of the proposed methodology opens up avenues for further exploration to understand and enhance the control methodologies for high DER-penetrated distribution networks. These include:
\begin{itemize}
    \item \textbf{PV addition prioritization:} The tools used in this work may be used by
    utilities to identify the sensitivity tolerance across their distribution networks enabling them to craft their renewables-tied incentives to ensure limited impacts to the system.
    \item \textbf{Elimination of SVRs, OLTCs, and capacitor banks:} While this work assumed fixed settings for SVRs, OLTCs, and capacitors, 
    there may be a path for future high DER-penetrated distribution networks to fully rely on inverter control and eliminate role of these components altogether.
    \item \textbf{Online algorithm implementation:} This work focused on improving the HC by implementing the most effective control algorithm. However, to ensure high-speed and real-time control, approaches for faster implementation including the use of ML may be pursued.  
\end{itemize}


%

\ifCLASSOPTIONcaptionsoff
  \newpage
\fi



%



{\footnotesize
\bibliographystyle{IEEEtran}
\bibliography{IEEEabrv, references}
}

%








\end{document}